\documentclass[conference]{IEEEtran}
\IEEEoverridecommandlockouts

\usepackage{cite}
\usepackage{amsmath,amssymb,amsfonts}
\usepackage{graphicx}
\usepackage{textcomp}
\usepackage{xcolor}
\usepackage{algorithm}
\usepackage{algpseudocode}
\usepackage{booktabs}
\usepackage{natbib}
\setcitestyle{numbers,open={[},close={]},comma,separate-uncertainty}

\usepackage{hyperref}
\hypersetup{
    colorlinks=true, 
    linkcolor=black, 
    citecolor=black, 
    filecolor=black,
    urlcolor=black 
}

\def\BibTeX{{\rm B\kern-.05em{\sc i\kern-.025em b}\kern-.08em
    T\kern-.1667em\lower.7ex\hbox{E}\kern-.125emX}}
\begin{document}

\title{BIPeC: A Combined Change-Point Analyzer to Identify Performance Regressions in Large-scale Database Systems\\
}

\author{\IEEEauthorblockN{Zhan Lyu$^\ast$}
\IEEEauthorblockA{\textit{SAP Labs China} \\
Xi'an, China \\
pal.lyu@sap.com}
\and
\IEEEauthorblockN{Thomas Bach$^\dagger$}
\IEEEauthorblockA{\textit{SAP} \\
Walldorf, Germany \\
thomas.bach03@sap.com}
\and
\IEEEauthorblockN{Yong Li$^\ddagger$}
\IEEEauthorblockA{\textit{SAP Labs China} \\
Xi'an, China \\
yong01.li@sap.com}
\and
\IEEEauthorblockN{Nguyen Minh Le$^\ddagger$}
\IEEEauthorblockA{\textit{SAP Labs China} \\
Xi'an, China \\
minh.le@sap.com}
\and
\IEEEauthorblockN{Lars Hoemke$^\ddagger$}
\IEEEauthorblockA{\textit{SAP} \\
Walldorf, Germany \\
lars.hoemke@sap.com}
}

\maketitle

\begin{abstract}
Performance testing in large-scale database systems like SAP HANA is a crucial yet labor-intensive task, involving extensive manual analysis of thousands of measurements, such as CPU time and elapsed time. Manual maintenance of these metrics is time-consuming and susceptible to human error, making early detection of performance regressions challenging. We address these issues by proposing an automated approach to detect performance regressions in such measurements. Our approach integrates Bayesian inference with the Pruned Exact Linear Time (PELT) algorithm, enhancing the detection of change points and performance regressions with high precision and efficiency compared to previous approaches. Our method minimizes false negatives and ensures SAP HANA's system's reliability and performance quality. The proposed solution can accelerate testing and contribute to more sustainable performance management practices in large-scale data management environments.
\end{abstract}


\section{Introduction}

In various domains such as medicine, aerospace, finance, business, meteorology, and entertainment, analyzing time series data—sequences capturing system behavior over time \cite{b1}—has become increasingly crucial. These time series are subject to shifts due to internal and external triggers, necessitating the detection of such abrupt changes, known as Change Point Detection (CPD) \cite{b2}. This field, while related to segmentation, edge, event, and anomaly detection \cite{b3}, focuses on identifying and modeling changes \cite{b4,b5,b6}. Over the decades, CPD's relevance has spanned multiple real-world applications, underscored by research in data mining, statistics, and computer science \cite{b7}.

In the context of SAP HANA \cite{b8,b9,b10}, performance testing is an essential part of the system's lifecycle, ensuring the robustness and efficiency that enterprises rely on. These systems form the backbone of modern data-driven decision-making processes, where rapid analysis of massive datasets is crucial. Performance testing, however, presents intricate challenges, demanding analysis of numerous performance metrics, such as CPU time and elapsed time. Traditionally, this has been a manual and labor-intensive process \cite{b11}, fraught with potential human error and the inefficiencies of prolonged testing cycles. Identifying performance regressions, or decreases in database performance after updates, is challenging due to the complexity of operations and subtle shifts in metrics. Manual practices consume resources, risking oversight and delays in SAP HANA deployment.

Our methodology aims to make a modest contribution to the existing body of knowledge by addressing a specific gap in performance regression detection, particularly in large-scale database systems like SAP HANA. Our approach ensures that genuine performance regressions are accurately identified while minimizing false negatives and unnecessary alarms. This robust detection framework reduces manual intervention and supports a more proactive and flexible system maintenance model.

Extensive testing across various datasets demonstrates that the proposed solution expedites testing enhances reliability, and ensures SAP HANA's performance quality.

Our list of contributions:
\begin{itemize}
\item We present and evaluate a new algorithm capable of identifying change points in time-series data of large-scale database systems.
\item We apply our approach to the performance analysis of SAP HANA, enhancing the degree of automation and reducing human resource investment.
\item We publish the source code for our approach.\footnote{https://doi.org/10.5281/zenodo.13319233}
\end{itemize}

The rest of this paper is organized as follows. Section \ref{2} presents two motivating examples to show why using natural language and execution information is necessary. Section \ref{3} presents some background knowledge used in our approach. Section \ref{4} presents our experiment design and dataset. Section \ref{5} reports our results. Section \ref{6} discusses some lessons learned. Section \ref{7} concludes this paper.

\subsection{Existing Performance Analysis and Limitations
}
Our performance testing infrastructure is developed on a robust continuous integration (CI) \cite{b12} system known as BMDB, which is central to our testing strategy. Benchmark Monitor (BM) \cite{b13} is a performance measurement framework and a web-based reporting application, particularly for performance regression tests. It helps to analyze the performance development of the SAP HANA database and supports the continuous integration process. 

In performance tests, measurements are considered the fundamental unit, capable of capturing metrics such as elapsed time, CPU time, memory consumption, and potentially others like I/O throughput or query metrics. These measurements are stored in a database, which fulfills several critical functions:
\begin{itemize}

\item\textbf{Performance Benchmarks: }The data allows for establishing performance benchmarks, against which current and future test results can be compared. This comparison is crucial for identifying performance improvements or regressions.

\item\textbf{Performance Trend Monitoring:} By storing historical data, it is possible to track the performance of systems over time. This longitudinal analysis can reveal trends, such as gradual improvements or the emergence of performance bottlenecks.

\item\textbf{Hardware Reliability Assessment:} The database can help identify malfunctioning or underperforming hardware by spotting inconsistent performance results, which may indicate "broken" or suboptimal hardware components.

\item\textbf{Test Quality and Stability Evaluation:} The collected data aids in assessing the reliability and stability of performance tests, ensuring that the tests are robust and produce consistent results.

\item\textbf{Benchmark Adaptation:} For performance improvement purposes, the system can automatically adjust the benchmark used for comparison. This adaptive approach ensures that benchmarks remain relevant and reflect current system performance, facilitating more accurate comparisons over time.

\end{itemize}

Performance testing in sophisticated systems like SAP HANA involves sifting through thousands of metrics, including CPU time and elapsed time. Additionally, there are numerous specialized measurements tailored to specific scenarios, including metrics for individual bug occurrences and their impact on system performance, measurements focused on particular types of queries or query patterns, and data associated with specific combinations of queries and system states, an effort traditionally maintained by hand, consuming considerable human labor, annotators need to spend at least 3 hours daily processing hundreds of time series, identifying change points, and opening tickets to developers. A paramount challenge within this realm is the detection of regressions in performance metrics. With the growing volume of performance test runs and the consequent expansion of time series data, the urgency to promptly detect regressions or improvements becomes crucial. This rapid identification is pivotal for maintaining software quality.

Our existing methods primarily involve manual comparisons of recorded performance metrics against results from new performance tests, such as response time, throughput, and resource utilization.  Given the extensive amount of data involved, this manual approach introduces considerable delays and has a low tolerance for errors.  The reliance on human analysis slows the process and increases the risk of overlooking subtle yet important performance deviations. 

Therefore, we need an advanced, automated method that can swiftly and reliably identify anomalies—be they regressions or improvements—allowing us to pinpoint and address potential issues rapidly. This transition to a more sophisticated system is essential for proactively and precisely maintaining our software's integrity and performance standards.

\section{Related Work}\label{2}

Upon reviewing various change-point detection techniques, it becomes evident that when employed for real-time analysis of extensive datasets, well-known methods such as CUSUM \cite{b14}, Binary Segmentation \cite{b15}, and classical PELT \cite{b16} often require more precision. Despite their extensive documentation, these methods struggle to handle massive datasets. This underscores the critical need for more advanced solutions capable of precisely scaling with the increasing volume and velocity of data in modern database systems.

\subsection{Traditional Change-Point Detection Methods}
\begin{table*}[ht]
\caption{Summary of change point detection methods.}

\centering
\scalebox{1}{
\begin{tabular}{lll}
\toprule
Name & Method & Reference \\
\midrule
AMOC & At Most One Change & Hinkley (1970) \\
BINSEG & Binary Segmentation & Scott and Knott (1974) \\
BOCPD & Bayesian Online Change Point Detection & Adams and MacKay (2007) \\
BOCPDMS & BOCPD with Model Selection & Knoblauch and Damoulas (2018) \\
CPNP & Nonparametric Change Point Detection & Haynes et al. (2017) \\
ECP & Energy Change Point & Matteson and James (2014) \\
KCPA & Kernel Change-Point Analysis & Harchaoui et al. (2009) \\
PELT & Pruned Exact Linear Time & Killick et al. (2012) \\
RBOCPDMS & Robust BOCPDMS & Knoblauch et al. (2018) \\
RFPOP & Robust Functional Pruning Optimal Partitioning & Fearnhead and Rigaill (2019) \\
SEGNEIGH & Segment Neighborhoods & Auger and Lawrence (1989) \\
WBS & Wild Binary Segmentation & Fryzlewicz (2014) \\
\bottomrule \\

\end{tabular}
}

\label{tab:change_point_methods}
\end{table*}
\begin{figure*}[htbp]
  \centering
  \includegraphics[width=0.8\linewidth]{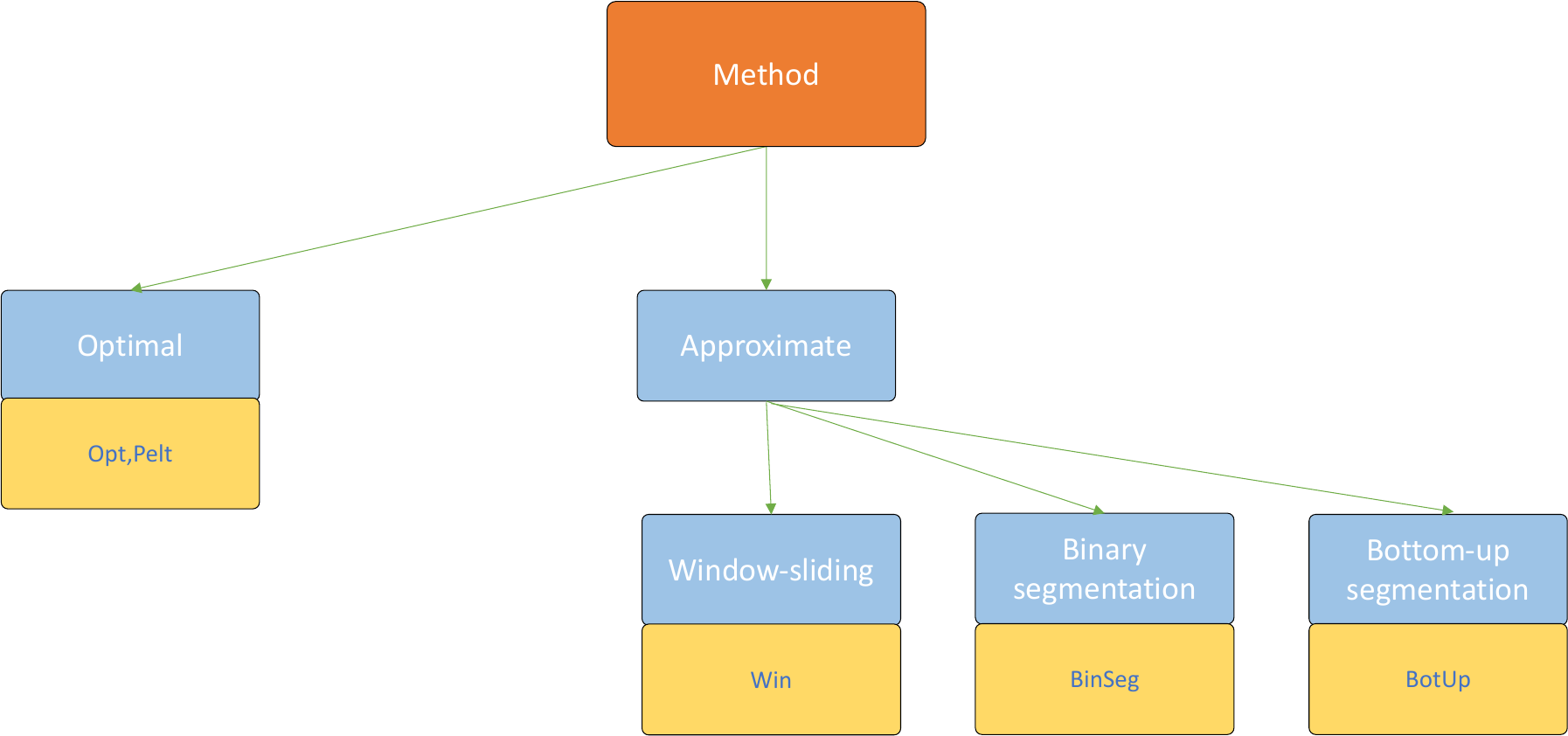}
  \caption{Typology of the methods described.}
  \label{fig:result}
\end{figure*}

\begin{figure*}[htbp]
  \centering
  \includegraphics[width=1\linewidth]{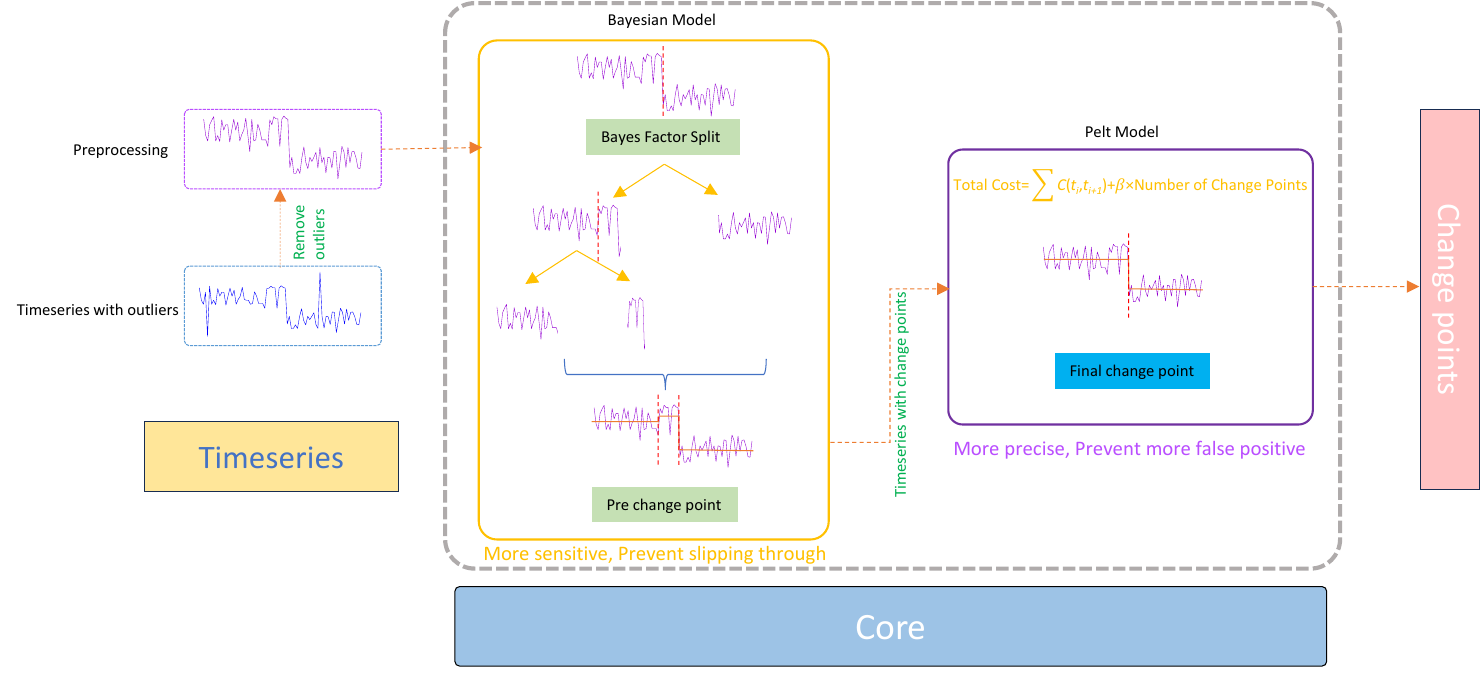}
  \caption{The architecture of BIPeC.}
  \label{fig:art}
\end{figure*}
Traditional change point detection broadly categorizes methodologies into optimal and approximate (see Fig. 1) \cite{b17}. Optimal methods, such as OptPelt \cite{b18}, exemplify efficiency, leveraging linear-time complexity and pruning techniques to expedite computation. These methods aspire to accurately determine the most probable set of change points by minimizing a cost function, Yet they come with the caveat of necessitating precise parameter calibration and an inherent sensitivity to their initial analytical setup.

Approximate methods, represented by the Window-sliding technique, provide a more intuitive approach. They are particularly well-suited to instances where change intervals are predefined, allowing for a straightforward implementation. However, their suitability wanes when faced with data complexity and acoustic clutter, where the exact location of change points is still being determined.  They are also known to impose substantial computational overhead.

Within approximate methods, Binary Segmentation stands out for its recursive efficiency, especially when the number of change points is undetermined. This method can briskly identify multiple shifts but may overlook closely spaced changes and struggle in noisy environments. On the other hand, Bottom-up Segmentation exhibits a propensity for detecting subtle changes, beginning with a wide net of potential change points and methodically merging segments. This method's strength in discerning small-scale shifts is countered by its vulnerability to noise and the potential for high computational load due to its initial detailed segmentation.

The shift from traditional to advanced methods has expanded change-point detection techniques, dealing with numerous challenges in large-scale, real-time performance analysis. The study by Smith \cite{b19} on Bayesian change-point inference laid the foundation for analyzing the probability of change points. Adams and MacKay \cite{b20} improved this method by introducing Bayesian online change-point detection, showcasing its adaptability and real-time application in different fields.

Traditional methods often face scalability and precision limitations when applied to complex data like those in SAP HANA performance testing. The binary segmentation technique, as developed by Scott and Knott \cite{b21}, and the non-parametric approaches by Haynes, Fearnhead, and Eckley \cite{b22}, and Matteson and James \cite{b23}, have contributed to the field by offering methods for grouping and detecting change points with improved computational efficiency and without stringent assumptions on data distribution.
Moreover, advancements by Killick, Fearnhead, and Eckley \cite{b24} in optimizing change-point detection algorithms highlight the shift towards reducing computational overhead, a crucial consideration for large-scale database systems like SAP HANA. The Wild Binary Segmentation (WBS) development by Fryzlewicz \cite{b25} represents a major advance in addressing multiple change-point detection with enhanced accuracy and computational feasibility.
The integration of robust Bayesian inference algorithms for non-stationary streaming data, as proposed by Knoblauch, Jewson, and Damoulas \cite{b26}, and the kernel-based change point analysis by Harchaoui, Moulines, and Bach \cite{b27}, reflect the ongoing innovation in the field, offering methods that are doubly robust and capable of handling complex data with high precision.

\subsection{Addressing Performance Regression Challenges in SAP HANA}

SAP HANA stands at the forefront of in-memory databases, distinguished for its rapid processing and accuracy. A fundamental component in safeguarding these hallmarks is the rigorous performance testing regime, which records many metrics, from CPU timings to memory usage. Traditionally, these metrics—or time-series data—are curated manually, demanding considerable time and human resources. One of the main obstacles in this process is identifying performance regression, which is an integral part of system assessment.

The prevalent method of static benchmarks to gauge performance outcomes shows its constraints, particularly in swiftly and precisely capturing any performance regression. A dynamic, data-driven methodology is required to discern patterns and anomalies in the time-series data autonomously. Such an algorithm would transform the performance testing landscape by infusing it with real-time, automatic change detection capacity, thereby liberating valuable resources and enhancing the overall efficiency and reliability of SAP HANA's performance evaluation.
Our literature review indicates a pronounced research gap in large-scale performance testing of large-scale database systems: real-time analytical methods must be revised in dealing with the complexity of SAP HANA's performance test time series. Traditional change-point detection algorithms often fail to identify shifts with high precision due to the intricate nature of such data streams. 

Our proposed algorithm, Bayesian Initiated PELT Confirmation (BIPeC), is precisely engineered to fill this gap. BIPeC marries the probabilistic strengths of Bayesian methods with the PELT algorithm's swift and precise change-point identification. The integration of these methodologies enables the BIPeC algorithm to excel in real-time analysis, improving accuracy in analyzing complex time series data, such as those found in SAP HANA performance tests.

The necessity of our approach stems from its precision in large-scale environments where traditional methods falter, 
our experiments show that BIPeC performs well with both complex and simpler time series data, making it a versatile tool for maintaining SAP HANA's performance quality and supporting sustainable performance management in large-scale data environments.

\section{IMPLEMENTATION}\label{3}

\subsection{Architecture}
Fig. 2 outlines the BIPeC framework, beginning with a preprocessing stage aimed at eliminating outliers and reducing noise, thus preparing the data for subsequent analysis. Following this, the data is introduced into the Bayesian model, which serves as the first critical phase of the framework. This model categorizes the time series into two groups: those exhibiting potential change points and those not. It marks the identified potential change points as pre-change points for further analysis.

The subsequent phase involves the PELT algorithm, designed to refine the detection process by optimizing the identification of actual change points from the pre-change points. This structured and sequential application of methodologies—from preprocessing to Bayesian analysis and culminating with the PELT refinement—offers substantial benefits. Firstly, the Bayesian model accelerates the data processing speed by swiftly pinpointing areas with potential changes. Secondly, the PELT algorithm builds upon this preliminary analysis by rigorously optimizing the pre-change points, ultimately leading to the discernment of final, authenticated change points.

This two-step approach expedites the algorithm's overall operation and assures that the outcomes are consistent and repeatable, even when the algorithm is applied multiple times to the same dataset. Consequently, the BIPeC framework is exceptionally adept at processing complex and extensive datasets, providing enhanced precision and reliability in change point detection. This makes BIPeC an essential tool for conducting high-quality, accurate analyses of large-scale data environments.

\subsection{Bayesian Change Point Detection Algorithm}
In employing the Bayesian approach to change point detection, a pivotal component is the Bayes factor, which quantitatively assesses the evidence for competing hypotheses regarding the presence of a change point. The Bayes factor is computed by contrasting the likelihoods of the observed data under two distinct hypotheses: \(H_1\), where the data is assumed to be homogeneous throughout (no change point), and \(H_2\), where the data is considered to have a change point that alters the underlying data distribution parameters.
\begin{itemize}

\item\textbf{Formulating the Hypotheses and Their Likelihoods:}
   For hypothesis, \(H_1\) (no change point), the likelihood of the data given this hypothesis, \(P(D|H_1)\), is computed as:
\begin{equation}
P(D|H_1) = \int p(D|\lambda) p(\lambda|H_1) d\lambda
\end{equation}

   Here, \(p(D|\lambda)\) represents the likelihood of the data \(D\) given the parameter \(\lambda\) (such as the mean of the Poisson or Gaussian distribution), and \(p(\lambda|H_1)\) is the prior distribution of \(\lambda\) under the assumption of no change.

   For hypothesis, \(H_2\) (existence of a change point), the likelihood, \(P(D|H_2)\), involves integration over parameters governing segments before (\(\lambda_1\)) and after (\(\lambda_2\)) the change point:

\begin{equation}
P(D|H_2) = \int\int p(D|\lambda_1, \lambda_2, t_s) p(\lambda_1, \lambda_2|H_2) d\lambda_1 d\lambda_2
\end{equation}
   where \(t_s\) denotes the time of the change point, \(p(D|\lambda_1, \lambda_2, t_s)\) is the likelihood of data given the parameters before and after the change, and \(p(\lambda_1, \lambda_2|H_2)\) is the joint prior distribution of these parameters.

\item\textbf{Bayes Factor Calculation:}
   The Bayes factor \(B\) is then calculated by taking the ratio of these two probabilities:
   \begin{equation}
B = \frac{P(D|H_2)}{P(D|H_1)}
\end{equation}
   This ratio effectively measures how much more likely the data is under the assumption of a change point than the assumption of no change. A high Bayes factor indicates strong evidence in favor of the existence of a change point.

\item\textbf{Practical Computation of Likelihoods:}
   In cases where data is assumed to follow a Poisson distribution, because, in Bayesian change point detection, the choice of the Poisson distribution is primarily because it aptly describes the count of independent events occurring within a fixed time or space interval, and its mean equals its variance, simplifying the data analysis process. The additive property of the Poisson distribution simplifies the analysis when merging multiple data sources, while its conjugate nature with Bayesian methods allows the posterior distribution to maintain an analytical form, reducing computational complexity. These characteristics make the Poisson distribution theoretically appropriate and highly effective in practical applications, especially when dealing with data that exhibits count characteristics. \(p(D|\lambda)\) for \(H_1\) might be expressed as the product of probabilities of observing each data point under a Poisson process with a constant rate \(\lambda\). Similarly, for \(H_2\), \(p(D|\lambda_1, \lambda_2, t_s)\) would be modeled as a product of Poisson probabilities with rate \(\lambda_1\) before \(t_s\) and \(\lambda_2\) after \(t_s\).

\item\textbf{Integration Over Parameters:}
   The integrals in the likelihood calculations generally require numerical methods, especially when the parameter space is ample or when the prior distributions are not conjugate to the likelihood functions. Methods like Markov Chain Monte Carlo (MCMC) \cite{b28} approximate these integrals for computational feasibility.

   \end{itemize}

This Bayesian formulation not only facilitates a rigorous assessment of the likelihood of change points in complex datasets but also allows the incorporation of prior knowledge into the analysis.

To enhance Bayesian change point detection for SAP HANA performance test time series data, we have tailored the algorithm by adjusting its parameters, such as threshold and window size, to better fit the characteristics of our datasets. Here is a summary of the critical attributes and modifications made.
\begin{itemize}

\item\textbf{Change Points Dictionary: } We have developed a dictionary of pandas data frames that maps each trajectory to specific attributes such as the time point split, log odds, start and end times, and the probability of change points. This structured approach allows for precisely identifying change points within our data.

\item\textbf{No Split Dictionary:} Another dictionary identifies time points where there is a noticeable peak in the probability of a split but which fall below a defined threshold. This feature is crucial for refining and validating potential change points, ensuring that only substantial changes are considered.

\item\textbf{State Emission Dictionary:} The state emission dictionary correlates each trajectory with the split segments' sample mean and standard deviation. This data is instrumental in generating a step function that effectively models the time series data changes.

\item\textbf{Log Gamma Function:} The inclusion of a log gamma function for all \( N \)s  enhances the computational efficiency of factorial-like calculations, which are pivotal in the statistical processes involved in the algorithm.

\item\textbf{Threshold Adjustment:} We have set a customizable log odds threshold for identifying splits. By lowering this threshold, the algorithm can be more sensitive to potential changes, allowing for a fine-tuned detection process tailored to SAP HANA performance data's specific noise and variability characteristics.

\item\textbf{Step Function Array:} A dictionary maps each trajectory to a numpy array representing the step function derived from the detected change points, providing a clear visual representation of changes over time.

\item\textbf{Refined Change Points:} When a refinement function is applied, the refined change point dictionary is populated with newly identified splits, allowing one to reevaluate and adjust initial findings based on more stringent criteria.

\item\textbf{Window Size and Stride:} The algorithm includes optional parameters for a moving window. By adjusting these parameters, the algorithm can analyze data within a flexible window frame, adapting to the dynamic nature of SAP HANA performance data without a fixed window unless specified.
\end{itemize}

The Bayesian change point detection algorithm has been customized to improve accuracy and relevance for performance monitoring and maintenance in our specific SAP HANA data environment.

The whole algorithm can be summarized as follows.

\begin{algorithm}
\caption{Bayesian Change Point Detection}
\begin{algorithmic}[1]
\Require Time series data \( Y \), distribution type $\text{dist}$
\Ensure List of  change points $\mathbf{CP}$
\Procedure{DetectChangePoints}{$\mathbf{Y}, \text{dist}$}
    \State $\text{Detector} \gets \text{new Detector}(\mathbf{Y}, \text{dist})$
    \State $\mathbf{CP} \gets \text{empty list}$
    \State $n \gets \text{length}(\mathbf{Y})$
    \For{$i \gets 1 \text{ to } n$}
        \State $\text{positions} \gets \text{empty list}$  \Comment{Track possible change points}
        \For{$t \gets 2 \text{ to length}(\mathbf{Y}[i])$}
            \State $\text{log\_odds} \gets \text{Detector.calculate\_bayes\_factor}(\mathbf{Y}[i], t)$
            \If{$\text{log\_odds} > \text{threshold}$}
                \State $\text{positions}.\text{append}(t)$  \Comment{Store position of change point}
            \EndIf
        \EndFor
        \If{$\text{positions} \neq \text{empty}$}
            \State $\mathbf{CP}.\text{extend}(\text{positions})$  \Comment{Append new change points}
        \EndIf
    \EndFor
    \State $\text{Detector.regenerate\_step\_function}()$
    \State \textbf{return} $\mathbf{CP}$
\EndProcedure
\end{algorithmic}
\end{algorithm}

\subsection{PELT Change Point Detection Algorithm}
PELT is an algorithm designed for change point detection. It optimizes finding change points by minimizing a cost function while applying a penalty for the number of change points to avoid overfitting \cite{b29}. The critical feature of PELT is its ability to perform in linear time complexity, making it highly efficient, especially for large datasets.
The PELT algorithm aims to minimize a cost function $C$ over the data $D$, segmented into $n$ segments, subject to a penalty $\beta$ for each additional change point added:

\begin{equation}
C(D, n) + \beta \times n
\end{equation}
Here, $C(D, n)$ represents the cost of segmenting the data $D$ into $n$ segments, and $\beta$ is the penalty term that controls the trade-off between the fit of the model and the number of change points. The RBF \cite{b30} model impacts how $C(D, n)$ is calculated by defining the expected change or similarity between points in different segments.
The RBF is a function in machine learning, often used in the context of Support Vector Machines (SVMs) \cite{b31} and interpolation. In change point detection, the RBF model specifies how the data is expected to change at each point, helping identify the change points more effectively. The RBF model measures the similarity or dissimilarity between different points in the data.
RBF implies that the PELT algorithm is applied with the Radial Basis Function to detect change points.

We optimized the penalty parameter through experiments with SAP HANA time series data, finding a \( \beta \) value that balances model fit and change points, thereby enhancing the accuracy and efficiency of large dataset analysis. This confirms the effectiveness of SAP HANA for high-performance time series analysis.

The whole algorithm can be summarized as follows.

\begin{algorithm}
\caption{PELT Change Point Detection}
\begin{algorithmic}[1] 
\State \textbf{Input:} Time series data \( Y \), penalty value \( \beta \)
\State \textbf{Output:} List of change points $\mathbf{CP}$
\Procedure{PELT}{}
    \State Initialize \( n \) as length of \( Y \)
    \State Initialize cost array \( F \) with \( F[0] = 0 \)
    \State Initialize change points list $\mathbf{CP}$ as empty
    \State Define \( R \) as set of active candidate change points, initially \( R = \{0\} \)

    \For{\( t \) from 1 to \( n \)}
        \State \( F[t] = \infty \)
        \State Initialize \( \text{last\_cp} = 0 \) (the last change point before \( t \))
        \For{each candidate \( s \) in \( R \)}
            \State Compute cost \( C \) from \( s \) to \( t \)
            \State \( \text{temp\_cost} = F[s] + C + \beta \)
            \If{\( \text{temp\_cost} < F[t] \)}
                \State \( F[t] = \text{temp\_cost} \)
                \State \( \text{last\_cp} = s \)
            \EndIf
        \EndFor
        \State Add \( t \) to \( R \)
        \State Use pruning to remove points from \( R \) that are no longer candidates

        \If{some stopping condition is met}
            \State \textbf{break}
        \EndIf
    \EndFor
    \State Backtrack from \( F[n] \) to determine all change points
    \State \textbf{return} $\mathbf{CP}$

\EndProcedure
\end{algorithmic}
\end{algorithm}

\subsection{Understanding BIPeC Approach}
\begin{figure}[htbp]
  \centering
  \includegraphics[width=1\linewidth]{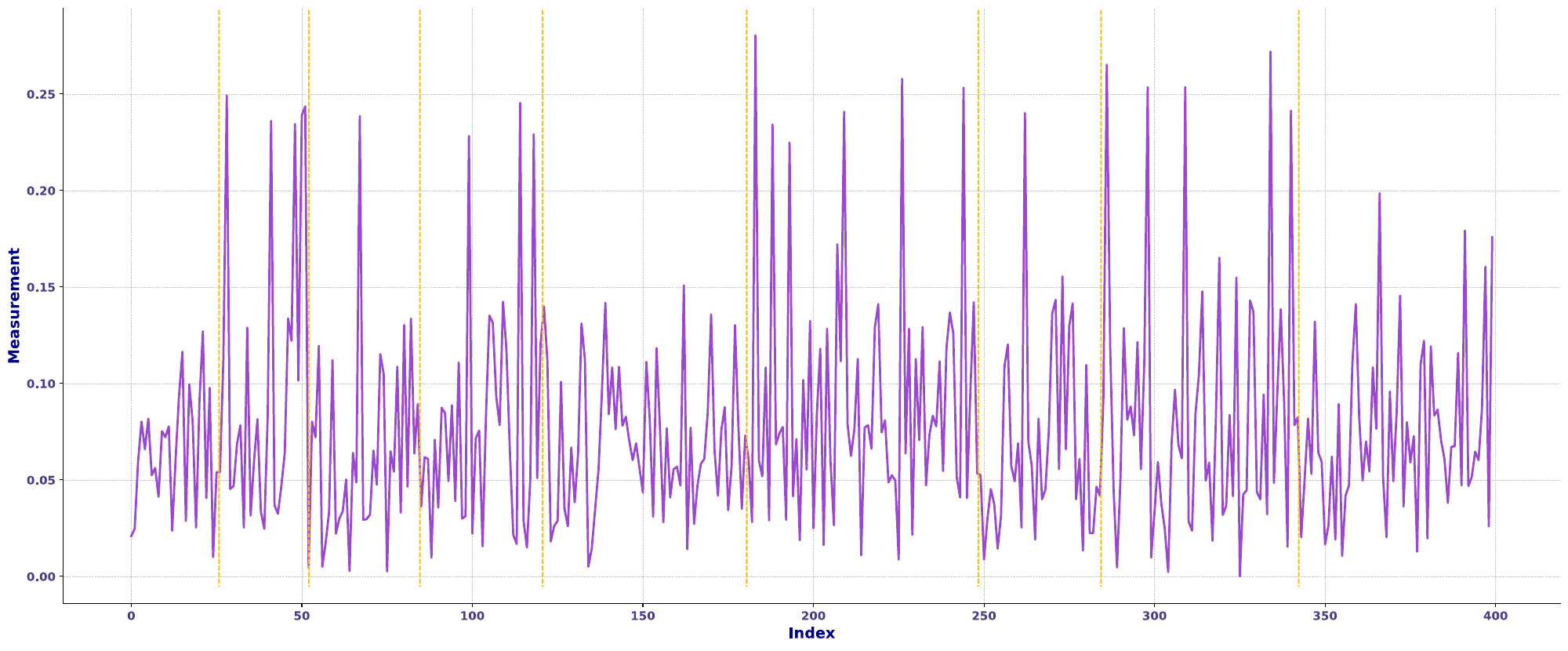}
  \caption{Bayesian algorithm's outcome.}
  \label{fig:fluctuation}
\end{figure}

\begin{figure}[htbp]
  \centering
  \includegraphics[width=1\linewidth]{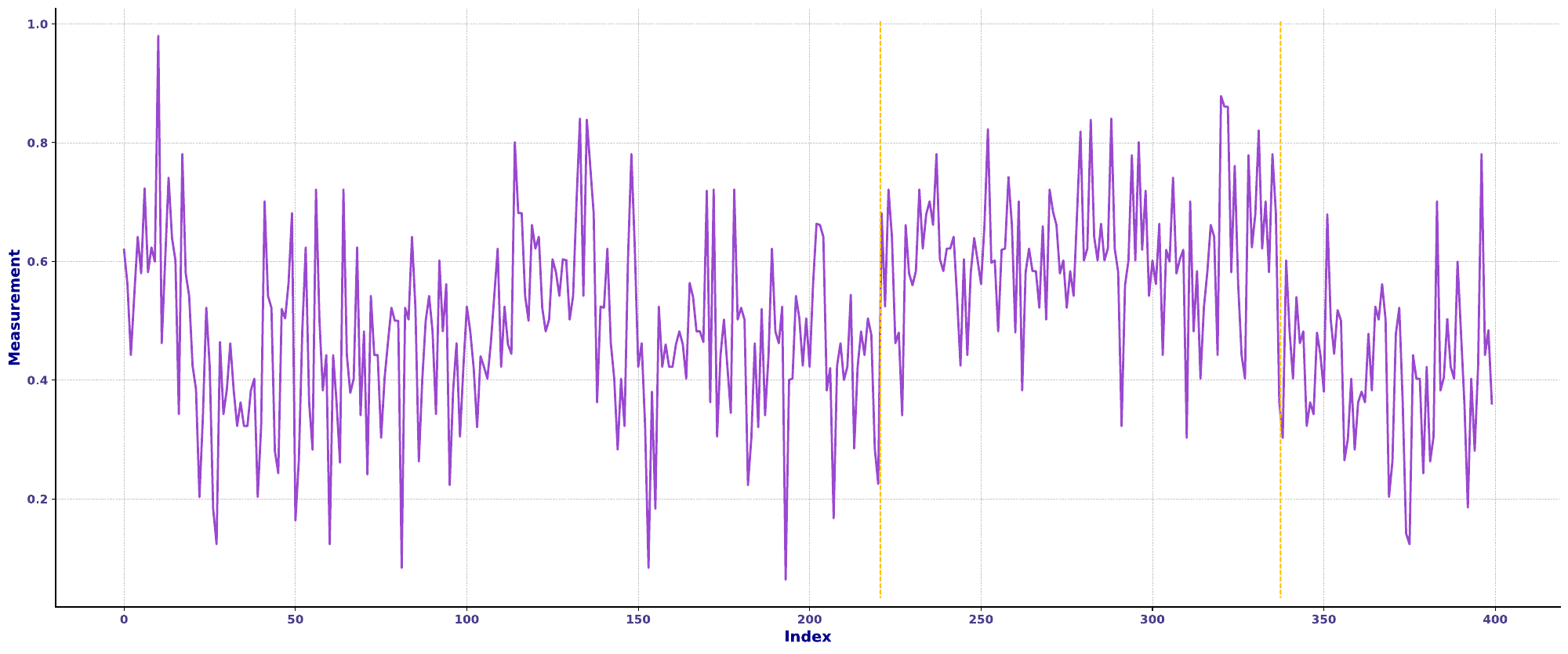}
  \caption{Pelt algorithm's outcome.}
  \label{fig:onlypelt}
\end{figure}
\begin{itemize}
        \item \textbf{Stepwise Strategy:}
Initiating the process with Bayesian methods for change point detection may be well-suited for addressing uncertainty and capturing global data characteristics without the need for excessive assumptions about model parameters. Subsequently, passing the change point detection results to the Pelt algorithm with RBF modeling allows for a more precise refinement and confirmation of the change point locations.
        \item \textbf{Diverse Change Point Detection Techniques:}
Bayesian methods and the Pelt algorithm have distinct change point detection strategies and strengths and weaknesses. Bayesian methods may be more sensitive to detecting potential change points but might lead to false positives, but the Pelt algorithm, with known models, may offer higher precision. Combining them can complement each other's weaknesses.
        \item \textbf{Reduced False Positives:}
By identifying potential change points using Bayesian methods and then employing the Pelt algorithm with RBF modeling for fine-tuning, we can lower the false positive rate and obtain more trustworthy results.
        \item \textbf{Noisy Data:}
If the data contains noise or random fluctuations, the Pelt algorithm may be sensitive to these fluctuations and detect false positives. Bayesian methods often introduce appropriate prior distributions to mitigate sensitivity to noise, reducing the occurrence of false positives.
        \item \textbf{Data Distribution:}
The characteristics of the data distribution can also influence performance. If the data distribution is regular, the Pelt algorithm may lead to false positives. Bayesian methods tend to adapt better to different data distributions.
        \item \textbf{Fluctuation Time Series:}
Fig. 3 illustrates the high sensitivity of the Bayesian approach, which, when applied to fluctuating time series, tends to identify multiple change points. This level of sensitivity often leads to the detection of more change points than anticipated, which, while not always aligning with our expectations, is a deliberate design choice to ensure rigorous analysis. Such stringency is crucial as it prevents the oversight of genuine change points, maintaining the integrity of our change point detection process.
        \item \textbf{Regular Time Series:}
Fig. 4 demonstrates that the PELT algorithm exhibits heightened sensitivity to regular time series, often identifying numerous change points. However, the Bayesian method processes such data more precisely, effectively discerning the valid change points. Consequently, a combined approach is necessary to integrate the strengths of both Bayesian and PELT methodologies. This synergy aims to mitigate the detection of spurious change points while ensuring no genuine ones are overlooked, thus achieving a balanced and accurate analysis.

\end{itemize}
\subsection{Parameter Adjustment and Evaluation}

Our BIPeC algorithm primarily utilizes the following parameters:

\begin{itemize}
    \item \textbf{pen:} The penalty value for the PELT model, which influences the algorithm's sensitivity to detecting change points by controlling the trade-off between detecting true changes and avoiding false positives.
    \item \textbf{window\_size:} The number of data points in each sliding window for analysis, determining the granularity of change-point detection and influencing how the algorithm captures short-term versus long-term changes in the time series.
    \item \textbf{chunk\_size:} The number of data points in each chunk for processing time series data, affecting the scope of data processed at a time and influencing the computational efficiency and resolution of the change-point detection.
    \item \textbf{log\_odds\_threshold:} The log threshold for Bayesian detection, controlling the decision boundary for detecting a change point based on the likelihood ratio, thereby influencing the balance between sensitivity and specificity in the detection process.
\end{itemize}

To ensure optimal performance, we conducted a systematic tuning process using hyperparameter optimization techniques. Specifically, we employed Hyperopt \cite{b32}, a popular library for hyperparameter optimization, to explore the parameter space and identify the best configurations for our algorithm. The optimization process was driven by maximizing the precision and F1 score of the detection results, ensuring that the parameters were tuned to achieve a balance between detecting true positives and minimizing false positives.

Traditional change-point detection methods involve tuning several key parameters, including:

\begin{itemize}
    \item \textbf{algorithm kernel:} Defines the type of kernel function used, influencing how changes in the data are modeled and detected.
    \item \textbf{alpha:} The significance level used in statistical tests, affecting the sensitivity of detection.
    \item \textbf{cost:} A penalty term that influences the cost function, balancing sensitivity to small changes and the risk of overfitting.
    \item \textbf{minsize:} The minimum segment size considered, preventing the detection of trivial changes.
    \item \textbf{threshold:} A cutoff value for detecting changes, with higher thresholds leading to fewer detections.
    \item \textbf{intensity:} Adjusts the weight given to certain features or data points, affecting the algorithm's sensitivity.
\end{itemize}

We also optimized these parameters using Hyperopt, with the goal of maximizing precision and F1 score. This approach ensured that each method was tested under its optimal conditions, providing a robust baseline for evaluating BIPeC’s performance.

The effectiveness of the parameter adjustments was evaluated through a series of experiments on datasets from SAP HANA performance tests and publicly available sources. We used precision and F1 score as the primary metrics for optimization, as these metrics provide a balanced view of detection performance by considering both true positives and false positives. The parameters optimized via Hyperopt allowed BIPeC to achieve a balance between detection accuracy and computational efficiency.

Because we conducted parameter optimization for both BIPeC and traditional methods on public datasets and SAP datasets, we ensured the comparison was as fair as possible. 

\begin{figure}[htbp]
  \centering
  \includegraphics[width=1\linewidth]{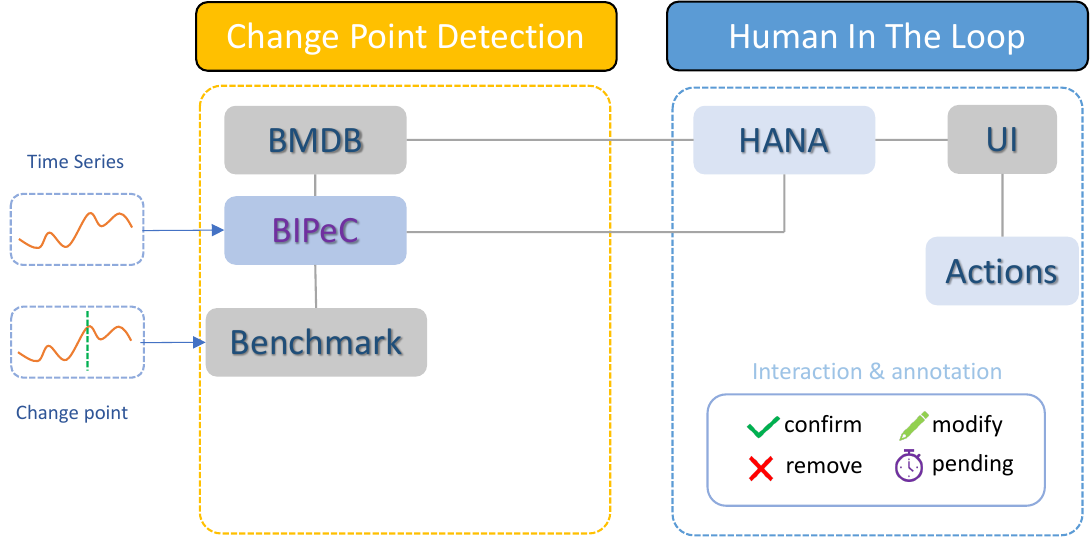}
  \caption{Feedback loop system.}
  \label{fig:loop}
\end{figure}
\subsection{Feedback Loop Service}
Our time series feedback service integrates automated change point detection with manual verification to enhance accuracy and efficiency in time series analysis. The process unfolds as follows:\\
\textbf{Automated Change Point Detection:} Initially, our tool automatically scans time series data to identify potential change points. These points represent substantial alterations in data trends, such as shifts in mean, variance, or other statistical properties.\\
\textbf{Manual Verification and Tagging:} Upon detection, these change points undergo manual scrutiny. Users engage in a review process where they can:
\begin{itemize}
\item \textbf{Confirm:} If a change point accurately represents a notable data event, it is marked as 'confirmed.'
\item \textbf{Remove:} In cases where a detected change point is deemed irrelevant or false, it is removed from consideration.
\item \textbf{Pending}: For ambiguous cases, where a clear decision isn't immediate, the change point is tagged as 'pending' for further review.
\item \textbf{Modify:} If the change point is valid but inaccurately positioned, users can manually adjust its position and mark it as 'modified.'
\end{itemize}
\textbf{Data Storage in BMDB Database:} Post-verification, all results, encompassing confirmed, removed, pending, and modified change points, are stored in our BMDB database. This centralized storage ensures data integrity and facilitates subsequent analysis.\\
\textbf{Model Parameter Optimization:} Leveraging this repository of manually verified results, our model periodically updates its parameters. This iterative refinement process utilizes real-world user feedback to enhance the model’s ability to identify and position change points in future datasets more precisely.\\
\textbf{Feedback Loop for Continuous Improvement:} The system is designed to foster a continuous feedback loop (see Fig. 5). User interactions and decisions feed directly into the system, driving ongoing enhancements in change point detection algorithms and overall system performance.

The feedback loop service enhances the precision of change point detection, making it essential for practical applications such as regression detection. By integrating automated detection with manual verification, it accurately identifies performance regressions, as demonstrated in environments like SAP HANA.

The manual verification step eliminates irrelevant change points, reducing false positives and enhancing reliability. The iterative feedback loop continuously fine-tunes model parameters, improving detection accuracy. This approach has been validated through testing on both public and SAP HANA datasets, demonstrating its effectiveness in complex, large-scale data environments

In summary, our time series feedback service is a dynamic tool that improves the quality and reliability of regression detection by effectively integrating automated processes with human expertise.

\section{Experiment Design and Dataset}\label{4}

We experimented with various combinations, including BOCPD and PELT, but found these methods fell short in performance and scalability compared to BIPeC, particularly in large-scale scenarios where BOCPD's computational complexity was less efficient. In contrast, BIPeC effectively integrates Bayesian methods with PELT, balancing real-time detection with efficiency, making it more suitable for large-scale environments like SAP HANA. We evaluated our approach using two distinct datasets: one with thousands of performance tests from SAP HANA instances and another from publicly available data. Both datasets underwent rigorous preprocessing to normalize metrics and ensure consistency. Our evaluation focused on detection accuracy, computational efficiency, and scalability. This comprehensive approach validated our methodology in the context of SAP HANA’s complex performance tests and confirmed its robustness across broader, public datasets. This dual-dataset strategy provided a thorough assessment of our methodology's performance across varied data landscapes.

\subsection{Pulic Dataset}
The public time series data for this study were sourced from a variety of online platforms \cite{b33}, including the World Bank, Eurostat, U.S. Census Bureau, GapMinder, and Wikipedia. Some of these series exhibited potential change points during global events like the 2007-2008 financial crisis, influencing variables such as Brent crude oil prices, U.S. business inventories, and GDP in several countries. Other series were linked to legislative actions, such as the UK's seat belt regulations and the Montreal Protocol's restrictions on CFC emissions. Additionally, we included datasets from previous studies on change point detection, such as the well-log, Nile River, and bee-waggle datasets. These datasets were selected based on their display of notable patterns, including sudden changes, seasonal effects, or the presence of outliers.

In total, we assembled 37 real-time series, consisting of 33 univariate and 4 multivariate series. Among these, one univariate series contained missing values, and several others exhibited seasonal trends. To further diversify the dataset, 5 synthetic "quality control" series with predetermined change points were discreetly added, bringing the total to 42 time series. The series varied in length, with an average of 328 data points, ranging from 15 to 991 points. 

The data set is made freely available to accelerate research on change point detection.\footnote{https://github.com/alan-turing-institute/TCPD}
\subsection{SAP HANA Performance Test Result Dataset}
Our SAP HANA performance test dataset is composed of time series data sourced from SAP HANA BMDB, specifically designed to capture a wide range of performance metrics that provide insight into the system’s behavior under various conditions. These metrics include detailed measurements related to individual bugs and their direct impact on system performance, as well as metrics focused on specific types of queries and query patterns. Additionally, the dataset includes data associated with particular combinations of queries and system states, offering a comprehensive view of performance under diverse scenarios. In total, we sourced 1,314 unidimensional time series.  varying in length, with the shortest being 269 points and the longest reaching 654 points. 76.33\% of these series have a length of exactly 400 points. All data originate from our database of performance test results, ensuring relevance to practical applications. To guarantee the accuracy of the annotations, we employed three different annotators for the labeling process. On average, the time series length in this dataset is approximately 420 points.

\subsection{Evaluation Metrics}
To assess the efficacy of BIPeC, our change point detection algorithm, we employ statistical tests on datasets with pre-established change points, treating the evaluation process akin to a task in machine learning. The core objective of BIPeC is to pinpoint change points within an input dataset with high precision, mirroring the process of deducing known outcomes from a set of training data. In our evaluation framework, we primarily leverage the F1 score as a holistic measure of the algorithm's accuracy, combining the insights of both precision and recall into a single metric. Alongside, we emphasize precision to directly gauge the algorithm's ability to identify true positives accurately without excess false positives. This dual metric approach, incorporating both F1 score and precision, ensures a balanced evaluation of BIPeC's performance. Through this refined assessment strategy, we aim to comprehensively evaluate BIPeC's capability to accurately detect change points, thus affirming its utility and effectiveness within complex data environments.

True Positives Definition
True positives are defined using two sets:

\(X^*\): the set of ground truth change points.

\(X\): the set of predicted change points.

True positives refer to the set of detected change points that correctly correspond to actual change points in the data. In other words, these are the change points that are both predicted by the model and present in the true set of change points. The accuracy of these predictions is assessed within a defined margin of error, denoted as \(M\). Specifically, if the difference between a predicted change point and the corresponding true change point is less than or equal to \(M\), then the predicted change point is considered correctly identified. This approach allows for some flexibility in the timing of the detected change points, accounting for small deviations. Mathematically, \(TP(X,X^*) : \{x \in X | \exists x^* \in X^* \text{ s.t. } |x - x^*| \leq M\}\).
 
 It is crucial to prevent double-counting a single \(X^*\) change point as matched by multiple predictions within the margin \(M\).

F1 score is selected for its robustness against variations in the size and density of the dataset and its capacity to penalize false positives while rewarding correct detections. It represents the harmonic mean of precision and recall, where:
\begin{equation}
F_1 = \frac{2 \cdot \text{precision} \cdot \text{recall}}{\text{precision} + \text{recall}}
\end{equation}
Precision is the fraction of correctly predicted positive observations from the total predicted positives.
\begin{equation}
\text{Precision} = \frac{|TP(X, X^*)|}{|X|}
\end{equation}Recall(or sensitivity) is the fraction of correctly predicted positive observations from all actual positives.
\begin{equation}
\text{Recall} = \frac{|TP(X, X^*)|}{|X^*|}
\end{equation}

\begin{figure}[htbp]
  \centering
  \includegraphics[width=1\linewidth]{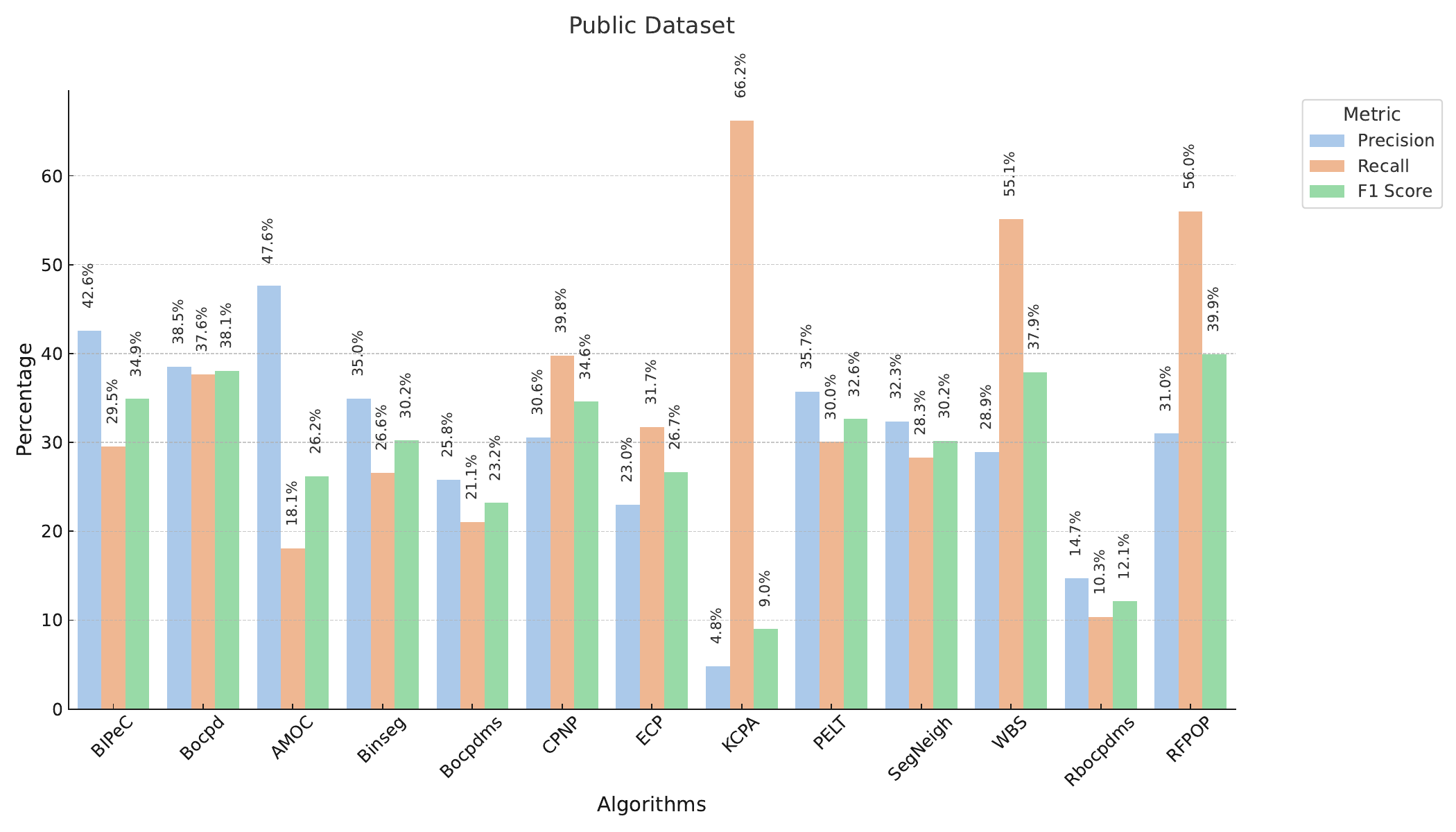}
  \caption{Results on public dataset.}
  \label{fig:result}
\end{figure}
\begin{figure}[htbp]
  \centering
  \includegraphics[width=1\linewidth]{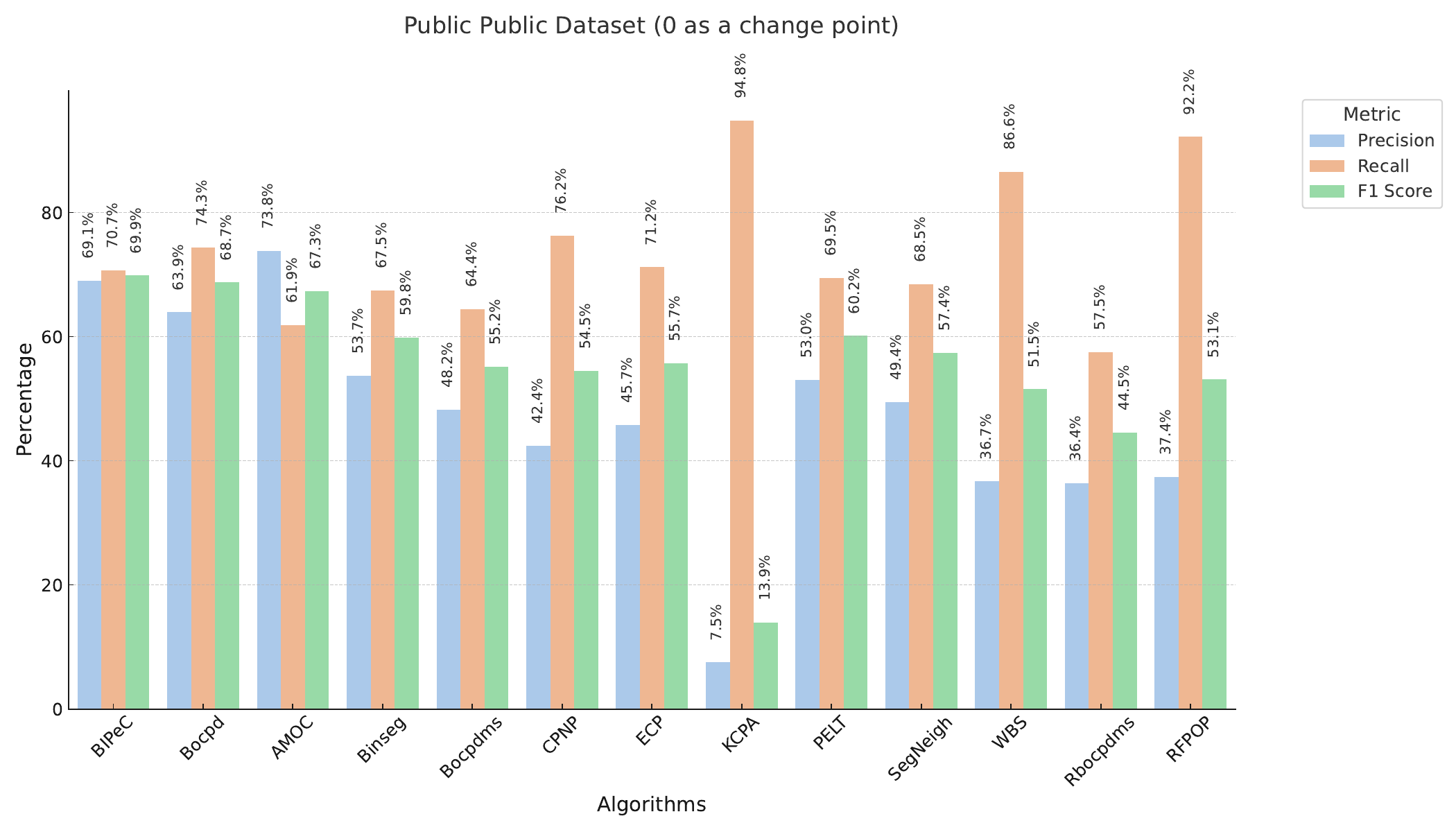}
  \caption{Results on public dataset with 0 as a default change point.}
  \label{fig:result}
\end{figure}
\begin{figure}[htbp]
  \centering
  \includegraphics[width=1\linewidth]{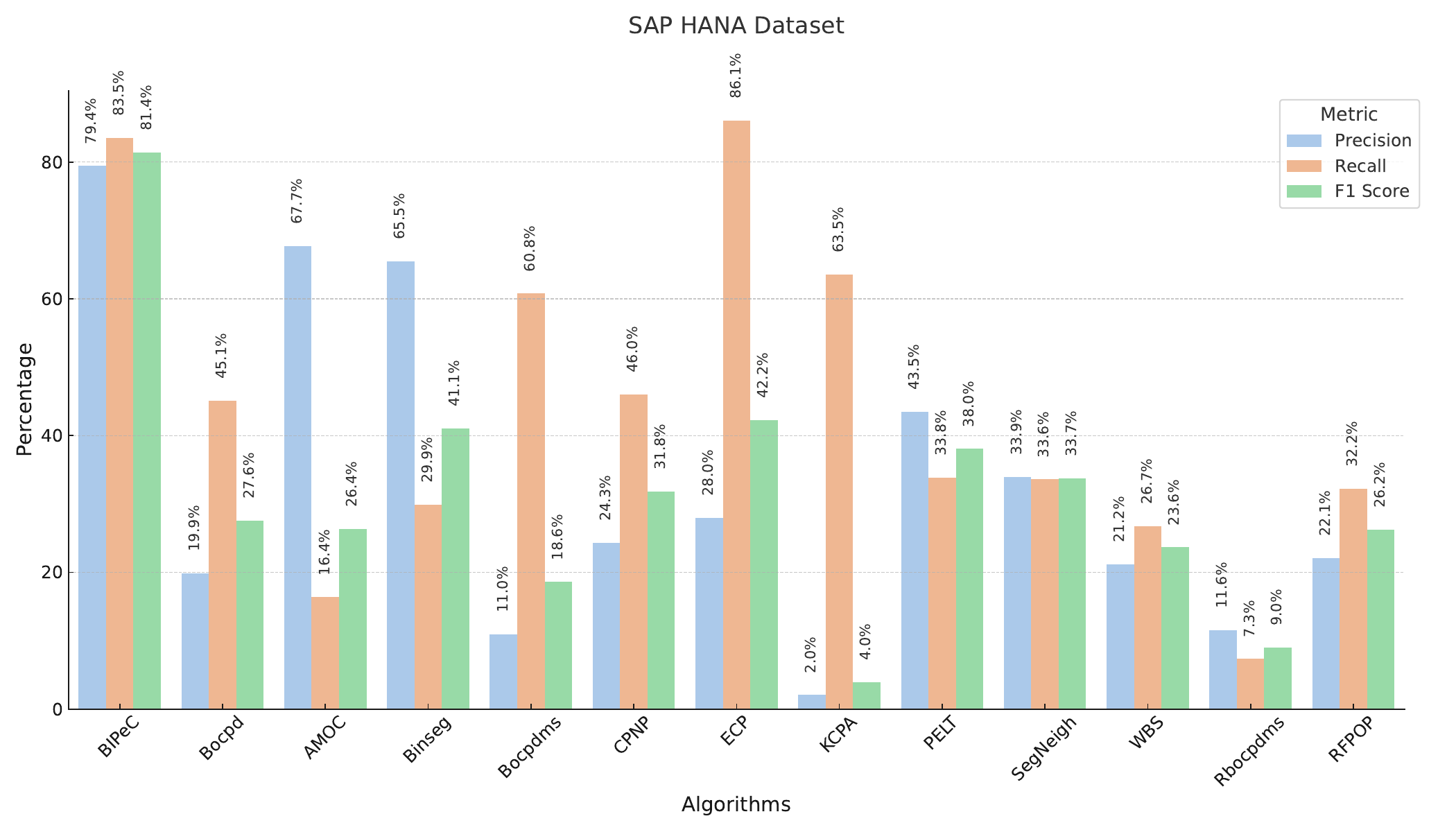}
  \caption{Results on SAP HANA dataset.}
  \label{fig:result}
\end{figure}
\begin{figure}[htbp]
  \centering
  \includegraphics[width=1\linewidth]{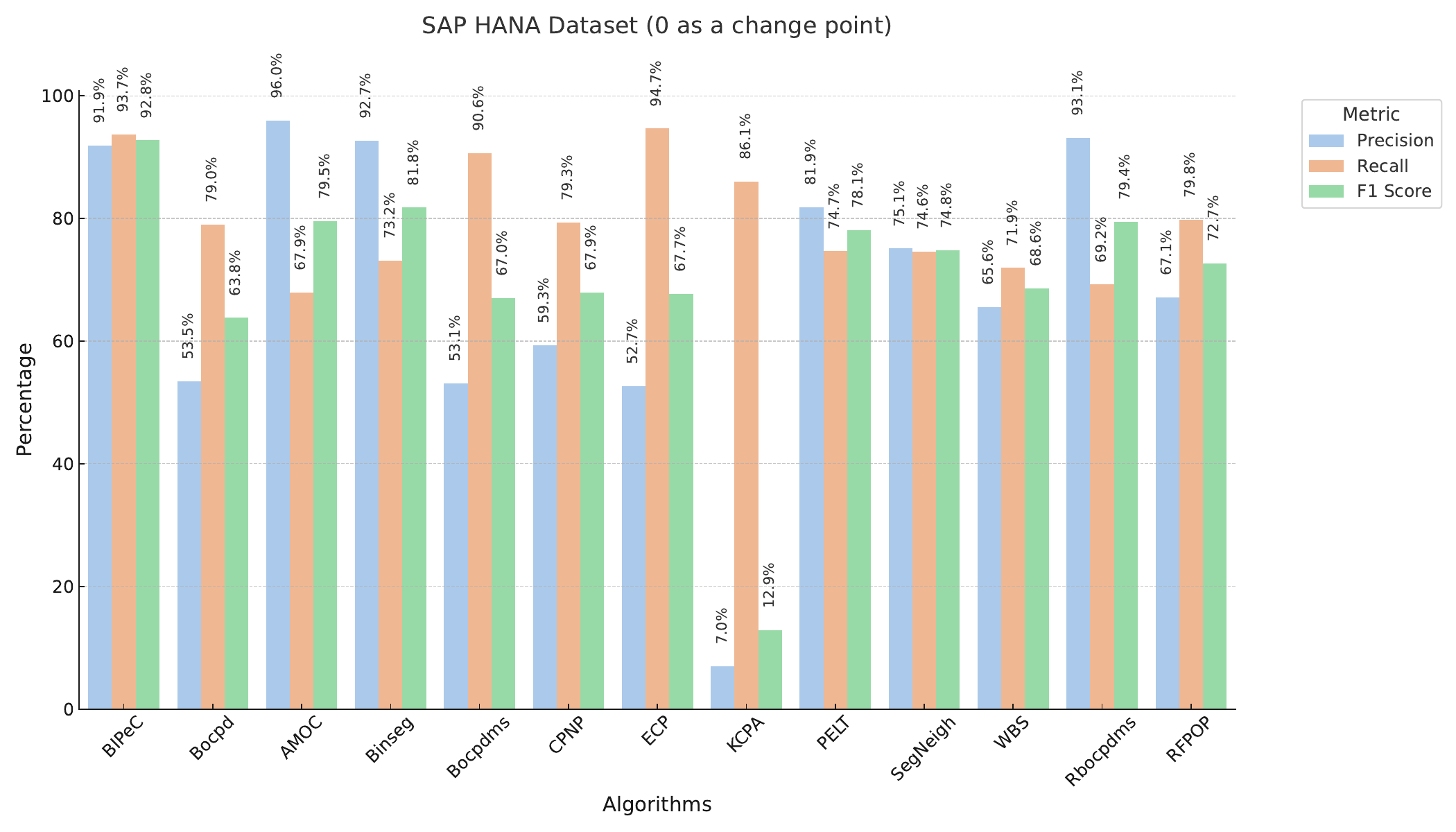}
  \caption{Results on SAP HANA dataset with 0 as a default change point.}
  \label{fig:result}
\end{figure}
\begin{figure}[htbp]
  \centering
  \includegraphics[width=1\linewidth]{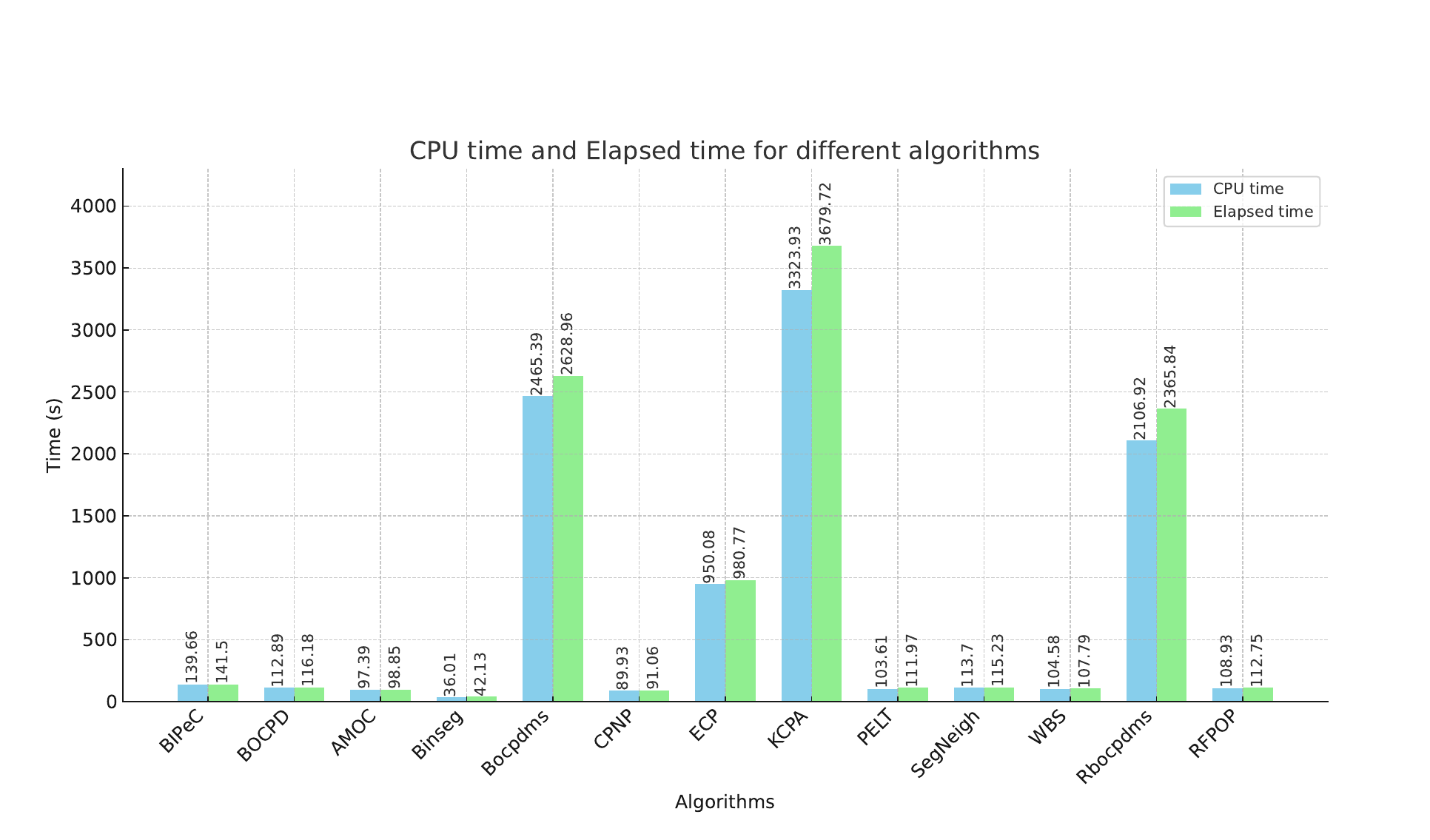}
  \caption{CPU time and elapsed time comparison on different algorithms.}
  \label{fig:result}
\end{figure}

\section{Results and Analysis}\label{5}

The results depicted in Fig. 7 and Fig. 9 showcase the comparative of various change point detection algorithms, including our BIPeC, which are obtained under the specific condition that zero is treated as the default change point. This means that both predictions and annotations within the datasets inherently recognize zero as a change point. 

In change point detection for time series data, treating the initial point (t = 0) as a default change point is a common practice. This approach helps in algorithmic initialization by providing a uniform reference point, ensuring consistency across different datasets and analyses. By marking t = 0 as a change point, the detection of subsequent changes becomes more straightforward, as the algorithm measures shifts relative to this baseline.

Additionally, including t = 0 helps manage boundary conditions, preventing distortions that might occur at the start of the series. This ensures a consistent framework for detecting and analyzing changes, making the process more reliable and reproducible. However, in cases where the initial conditions are not representative of the series' overall behavior, excluding t = 0 as a change point might be more appropriate, allowing the methodology to adapt to specific analysis needs.

For BIPeC, even under these constraints, the algorithm performs excellently on both the public and BMDB datasets. On the public dataset, BIPeC has an F1 score of about 70\% and a precision of about 69\%. In contrast, on the BMDB dataset, BIPeC has an F1 score of about 93\% and a precision of about 92\%, which is the first overall performance, showing that it is good at identifying other valid change points in the data while correctly handling the imposed default conditions.BIPeC's performance demonstrates its high detection accuracy and resistance to potential biases introduced by the default zero change point assumption.

It is worth noting that BIPeC's effectiveness relative to other algorithms, such as BINSEG, PELT, and BOCPD, could partly be attributed to its algorithmic architecture, which possibly incorporates sophisticated mechanisms to mitigate the influence of default change point assumptions on the overall detection process. This nuanced capability allows BIPeC to deliver reliable results even when the datasets are pre-conditioned with certain expectations, reinforcing the algorithm's robustness and adaptability in varied analytical contexts.

Fig. 6 and Fig. 8 show our results without restricting zero as the default change point. On the public dataset, BIPeC's precision and F1 score are still good, but on the SAP HANA BMDB dataset, BIPeC's precision is about 79\% and F1 score is about 81\%, and the F1 score of other algorithms is below 50\%, which shows that our BIPeC algorithm outperforms all others in Large-scale Database Systems, and this is precisely what we need, we only care about the valuable change point in the business scenario.

Without assuming zero as a change point, BIPeC's capacity to identify shifts within the data is revealed, underscoring its robustness and precision. The graph reflects BIPeC’s sophisticated design, which likely allows it to effectively manage and analyze the complexity inherent in performance test data, leading to high precision.

The two conditions present a comprehensive picture of BIPeC's capabilities. Under default change point assumption conditions, BIPeC demonstrated high precision, but its performance is even more remarkable without this assumption. This suggests that BIPeC's algorithm is more than just fine-tuned to handle the artificial boost of treating zero as a change point but is genuinely adept at discerning the nuances of change point data. Fig. 10 shows the actual performance of BIPeC and other algorithms on all datasets. Overall, although the resource consumption of BIPeC is not optimal, it is acceptable, and these tests were experimented on all datasets and run 10 times to take the average value. When combined, these insights substantiate the superiority of BIPeC in change point detection, particularly within the complex data landscapes of performance testing. The algorithm's adaptability to both conditions indicates its potential to provide reliable, nuanced analysis, proving an indispensable tool in performance monitoring and evaluation.

\section{Lessons Learned}\label{6}

Initially, our dataset only covered half a year, and we conducted daily performance tests, which resulted in many data points. This extensive data allowed us to better handle inaccuracies in the results, which led to a noticeable decrease in false positive detections of change points. We intend to conduct experiments with datasets spanning more extended periods to determine if we can further reduce the false positive rate.

This systematic approach to evaluation, combined with the insights gained from the extended application, has established BIPeC as the preferred solution for accurately identifying meaningful data changes.

In change point detection, a larger dataset is crucial. More data points provide richer information, enabling algorithms to better identify patterns and trends, improving detection accuracy. A large dataset helps distinguish between noise and actual signals, reducing false positives and false negatives. It also enhances model stability by allowing more accurate parameter estimation. Additionally, larger datasets can reveal subtle and complex changes, making detection methods more adaptable and flexible. They also help analyze long-term trends and cyclical variations, offering valuable insights for strategic decision-making. Overall, more data improves the accuracy, reliability, and effectiveness of change point detection.

Our experiments have shown that our algorithm remains effective even with smaller datasets. While the algorithm can operate with limited data, the effectiveness depends on the complexity and variability of the data. We recommend having a sufficient number of data points to capture the essential dynamics, which ensures reliable detection. Although our algorithm can provide reliable insights with smaller datasets, a larger dataset is preferable as it generally yields more accurate results. Proper calibration and adjustment of detection parameters are crucial to maximize effectiveness, enhancing operational efficiency and decision-making processes.

\section{Conclusion}\label{7}

This paper investigated a method for identifying changes in performance metrics in large database systems. Combining a Bayesian Probabilistic Model with the PELT algorithm created a highly effective approach for real-time performance analysis. The Bayesian-PELT framework is exact and practical for accurately detecting performance shifts. This method improves the accuracy of performance monitoring and paves the way for future advancements in the field, influencing theoretical exploration and real-world applications in large-scale databases.


\begin{thebibliography}{00}
\bibitem{b1} Kacprzyk, J., Wilbik, A., \& Zadroiny, S. (2006, October). Capturing the essence of a dynamic behavior of sequences of numerical data using elements of a quasi-natural language. In 2006 IEEE International Conference on Systems, Man and Cybernetics (Vol. 4, pp. 3365-3370). IEEE.
\bibitem{b2} Kawahara, Y., \& Sugiyama, M. (2009, April). Change-point detection in time-series data by direct density-ratio estimation. In Proceedings of the 2009 SIAM international conference on data mining (pp. 389-400). Society for Industrial and Applied Mathematics.
\bibitem{b3} Chandola, V., Banerjee, A., \& Kumar, V. (2009). Anomaly detection: A survey. ACM computing surveys (CSUR), 41(3), 1-58.
\bibitem{b4}  Boettcher, M. (2011). Contrast and change mining. Wiley Interdisciplinary Reviews: Data Mining and Knowledge Discovery, 1(3), 215-230.
\bibitem{b5} Hido, S., Idé, T., Kashima, H., Kubo, H., \& Matsuzawa, H. (2008). Unsupervised change analysis using supervised learning. In Advances in Knowledge Discovery and Data Mining: 12th Pacific-Asia Conference, PAKDD 2008 Osaka, Japan, May 20-23, 2008 Proceedings 12 (pp. 148-159). Springer Berlin Heidelberg.
\bibitem{b6} Scholz, M., \& Klinkenberg, R. (2007). Boosting classifiers for drifting concepts. Intelligent Data Analysis, 11(1), 3-28.
\bibitem{b7} Montanez, G., Amizadeh, S., \& Laptev, N. (2015, February). Inertial hidden markov models: Modeling change in multivariate time series. In Proceedings of the AAAI Conference on Artificial Intelligence (Vol. 29, No. 1).
\bibitem{b8} Färber, F., Cha, S. K., Primsch, J., Bornhövd, C., Sigg, S., \& Lehner, W. (2012). SAP HANA database: data management for modern business applications. ACM Sigmod Record, 40(4), 45-51.
\bibitem{b9} Färber, F., May, N., Lehner, W., Große, P., Müller, I., Rauhe, H., \& Dees, J. (2012). The SAP HANA Database--An Architecture Overview. IEEE Data Eng. Bull., 35(1), 28-33.
\bibitem{b10} May, N., B\"{o}hm, A., \& Lehner, W. (2017). Sap hana–the evolution of an in-memory dbms from pure olap processing towards mixed workloads.
\bibitem{b11} Labor Intensive Definition \& Example.
\bibitem{b12} Rehmann, K. T., Boehm, A., Lee, D. H., \& Wiemers, J. (2014, June). Continuous performance testing for SAP HANA. In Proceedings of the First International Workshop on Reliable Data Services and Systems (RDSS). Snowbird, UT, USA.
\bibitem{b13} Rehmann, K. T., Seo, C., Hwang, D., Truong, B. T., Boehm, A., \& Lee, D. H. (2016). Performance monitoring in sap hana's continuous integration process. ACM SIGMETRICS Performance Evaluation Review, 43(4), 43-52.
\bibitem{b14} Basseville, M., \& Nikiforov, I. V. (1993). Detection of abrupt changes: theory and application (Vol. 104). Englewood Cliffs: Prentice hall.
\bibitem{b15} Fryzlewicz, P. (2014). Wild binary segmentation for multiple change-point detection.
\bibitem{b16} Wambui, G. D., Waititu, G. A., \& Wanjoya, A. (2015). The power of the pruned exact linear time (PELT) test in multiple changepoint detection. American Journal of Theoretical and Applied Statistics, 4(6), 581.
\bibitem{b17} Truong, C., Oudre, L., \& Vayatis, N. (2020). Selective review of offline change point detection methods. Signal Processing, 167, 107299.
\bibitem{b18} Baxter, R. A., \& Oliver, J. J. (1994). MDL and MML: Similarities and differences. Dept. Comput. Sci. Monash Univ., Clayton, Victoria, Australia, Tech. Rep, 207.
\bibitem{b19} Smith, A. F. (1975). A Bayesian approach to inference about a change-point in a sequence of random variables. Biometrika, 62(2), 407-416.
\bibitem{b20} Adams, R. P., \& MacKay, D. J. (2007). Bayesian online changepoint detection. arXiv preprint arXiv:0710.3742.
\bibitem{b21} Scott, A. J., \& Knott, M. (1974). A cluster analysis method for grouping means in the analysis of variance. Biometrics, 507-512.
\bibitem{b22} Haynes, K., Fearnhead, P., \& Eckley, I. A. (2017). A computationally efficient nonparametric approach for changepoint detection. Statistics and computing, 27, 1293-1305.
\bibitem{b23} Matteson, D. S., \& James, N. A. (2014). A nonparametric approach for multiple change point analysis of multivariate data. Journal of the American Statistical Association, 109(505), 334-345.
\bibitem{b24} Killick, R., Fearnhead, P., \& Eckley, I. A. (2012). Optimal detection of changepoints with a linear computational cost. Journal of the American Statistical Association, 107(500), 1590-1598.
\bibitem{b25} Fryzlewicz, P. (2014). Wild binary segmentation for multiple change-point detection.
\bibitem{b26} Knoblauch, J., Jewson, J. E., \& Damoulas, T. (2018). Doubly Robust Bayesian Inference for Non-Stationary Streaming Data with $\beta $-Divergences. Advances in Neural Information Processing Systems, 31.
\bibitem{b27} Harchaoui, Z., Moulines, E., \& Bach, F. (2008). Kernel change-point analysis. Advances in neural information processing systems, 21.
\bibitem{b28} Gilks, W. R., Richardson, S., \& Spiegelhalter, D. (Eds.). (1995). Markov chain Monte Carlo in practice. CRC press.
\bibitem{b29} Hawkins, D. M. (2004). The problem of overfitting. Journal of chemical information and computer sciences, 44(1), 1-12.
\bibitem{b30} Sun, T. Y., Liu, C. C., Lin, C. L., Hsieh, S. T., \& Huang, C. S. (2009). A radial basis function neural network with adaptive structure via particle swarm optimization (pp. 423-436). IntechOpen.
\bibitem{b31} Chapelle, O., Vapnik, V., Bousquet, O., \& Mukherjee, S. (2002). Choosing multiple parameters for support vector machines. Machine learning, 46, 131-159.
\bibitem{b32} Bergstra, J., Yamins, D., \& Cox, D. (2013, February). Making a science of model search: Hyperparameter optimization in hundreds of dimensions for vision architectures. In International conference on machine learning (pp. 115-123). PMLR.
\bibitem{b33} Van den Burg, G. J., \& Williams, C. K. (2020). An evaluation of change point detection algorithms. arXiv preprint arXiv:2003.06222.

\end{thebibliography}
\end{document}